\documentclass[aps,12pt,showkeys,nofootinbib]{revtex4}

\newcommand{\be}{\begin{eqnarray}}
\newcommand{\ee}{\end{eqnarray}}
\newcommand{\ba}{\begin{array}}
\newcommand{\ea}{\end{array}}
\newcommand{\ds}{\displaystyle}
\newcommand{\dpa}{\partial}
\newcommand{\pa}[1]{\left(#1\right)}
\newcommand{\paq}[1]{\left[#1\right]}
\newcommand{\pag}[1]{\left\{#1\right\}}

\usepackage[english]{babel}
\usepackage{amsmath}
\usepackage{amssymb}
\usepackage{comment}
\usepackage{graphicx}

\begin{document}

\title{A Gravitational non-Radiative Memory Effect}

\author{Hebertt Leandro}
\affiliation{Departamento de F\'\i sica Te\'orica e Experimental, Universidade Federal do Rio Grande do Norte, Natal-RN 59078-970, Brazil}

\author{Riccardo Sturani}
\affiliation{International Institute of Physics, Universidade Federal do Rio Grande do Norte, Campus Universit\'ario, Lagoa Nova, Natal-RN 59078-970, Brazil}

\email{hlsilva@fisica.ufrn.br, riccardo@iip.ufrn.br}

\begin{abstract}
  We revisit the issue of \emph{memory effects}, i.e. effects giving rise to
  a net cumulative change of the configuration of test particles, using a
  toy model describing the emission of
  radiation by a compact source and focusing on the \emph{scalar},
  hence non-radiative, part of the Riemann curvature.
  Motivated by the well known fact that gravitational radiation is accompanied by a \emph{memory}
  effect, i.e. a permanent displacement of the relative separation of test particles,
  present after radiation has passed, 
  we investigate the existence of an analog effect in the \emph{non-radiative} part of
  the gravitational field.
  While quadrupole and higher multipoles undergo oscillations responsible
  for gravitational radiation, energy, momentum and angular momentum are
  conserved charges undergoing non-oscillatory change due to radiation emission.
  We show how the source re-arrangement due to radiation emission produce
  time-dependent scalar potentials which induce a time variation in the scalar
  part of the Riemann curvature tensor.
  As a result, on general grounds a \emph{velocity memory} effect appears,
  depending on the inverse of the \emph{square} of the
  distance of the observer from the source, thus making it almost impossible
  to observe, as shown by comparison to the planned gravitational detector
  noise spectral densities.
\end{abstract}

\keywords{Gravitational potential, memory effect, General Relativity}

\maketitle

\section{Introduction}

The recent detections of gravitational waves (GWs) \cite{LIGOScientific:2018mvr,Abbott:2020niy},
beside marking the beginning of the new science dubbed GW Astronomy, have
triggered scientific interest over all aspects of GW production and detection.

The gravitational field sourced by an astrophysical compact object, with
mass and velocity multipoles changing with time, includes both a radiative and
a non-radiative (or \emph{longitudinal}) part, the former being the gravitational radiation
with $1/r$ behavior (being $r$ the distance between source and
observer), the latter being sourced by conserved charges, mass and angular
momentum, which in the static limit enters the Riemann curvature
tensor components as $1/r^3$ (in the standard 3+1 dimensional space).
Gravitational radiation carries the fingerprints of source, its frequency
evolution being determined by the internal dynamics of the source constituents.
Gravitational non-radiative modes undergo secular changes as energy, momentum
and angular momentum is dragged out of the system by radiation.

The focus of the present work is to study the \emph{non-radiative} part of the gravitational
field and related observational effects which manifest as a different status of
test particles between before and after the passage of radiation,
thus producing a cumulative change in suitable observables which does not
vanish at late times.

The presence of \emph{memory} effects was first noticed in linearised gravity
already in \cite{1974SvA....18...17Z}
and then first analysed in \cite{Braginsky:1986ia} using geodesic deviation
equation, where the passage of GWs
sourced by moving massing objects was identified to cause a \emph{permanent}
displacement between test particles, not fading away after the gravitational
perturbation as gone quiet

The standard observable to be considered is the geodesic deviation,
entering a dynamical equation which can be expressed in terms of space vector with components $D^i$
parameterising the separation between nearby geodesics and the electric
part of the Riemann tensor
\be
\label{eq:geo_dev}
\ddot D^i=R^i_{\ 0j0}D^j\,,
\ee
an overdot denoting derivative with respect to time, taken to be the $0$-th coordinate.
The \emph{tensorial} contribution to eq.~(\ref{eq:geo_dev}) contains the effect
of gravitational waves, including the memory effect described in
\cite{Thorne:1992sdb,Wiseman:1991ss} consisting in 
two test particles acquiring a permanent displacement (and no relative velocity)
after the radiation has passed through them.
As the tensor part of the Riemann tensor in eq.(\ref{eq:geo_dev}) is proportional
to $\ddot h^{(TT)}_{ij}$, where $h^{(TT)}_{ij}$ is the transverce-traceless (TT) part of
the metric perturbation, when its asymptotic values at $t\to -\infty$ and $t\to \infty$ differ, one has
$$
\Delta D^i\propto\lim_{t\to\infty}\pa{h^{(TT)}(t)_{ij}-h^{(TT)}_{ij}(-t)}D^j\neq 0\,,
$$
hence the test particle will be at rest and will carry a ``memory'' of the passage
of radiation long after it is gone, even if local physics experiments will show no deviation
from Minkowski space.

The non-TT part, i.e.~the longitudinal, non-radiative part of the gravitational
field contains a $1/r$ Coulomb type gravitational field that cannot change because
of energy-momentum conservation, however we will show that it contains a
$1/r^2$ part that undergoes an impulsive change at the passage of a burst of
radiation and which can leave in principle a signature in a detector,
as it imparts a \emph{velocity kick}, i.e. test particles acquire relative velocity
after burst passage.

In literature it is usual to distinguish between the linear and a
non-linear \emph{Christodoulou} effect, due to the seminal analysis of
\cite{Christodoulou:1991cr},
where the energy momentum tensor of GW radiation is considered as
an additional source of GW radiation: being the energy momentum tensor quadratic
in GW amplitude this is a non-linear effect. Another peculiarity of
the non-linear Christodoulou memory effect is that the source is neither slowly
moving (being made of radiation it moves at the speed of light) nor confined to
a small region of space (the energy momentum tensor of GWs falling off as
$1/r^2$). 
Actually an energy momentum tensor made of radiation as a source of GWs was
already studied in \cite{1978ApJ...223.1037E,Turner:1978jj} (see also 
the more recent \cite{1997A&A...317..140M}), in the context of anisotropic
neutrino emission by supernovae
(at the time neutrinos were compatible with having zero mass) leading as well 
to memory terms.

Memory effects can also be considered as part of a wider category of \emph{hereditary} effects, so named as their value at any
instant of time depends on the history of the source rather than source
status at retarded time. The term \emph{hereditary} has been used
to indicate the correlation in the dynamics of a gravitating system between
arbitrarily large time spans for the first time in \cite{Blanchet:1987wq}.
There, within the context of the post-Newtonian approximation
to General Relativity, the \emph{tail} effect onto the metric due to scattering
of gravitational radiation with the background curved by the same source
emitting GWs were studied. Subsequently \cite{Blanchet:1992br} showed
another type of hereditary effect due to
the radiation of GWs sourced by the stress-energy tensor of GWs, causing a
cumulative change in the waveform, i.e. a \emph{memory} effect that does not
vanish after the passage of the radiation.
\footnote{Whether this neutrino radiation is actually of
memory or tail type is discussed in ch. 10.5.4 of \cite{maggiore2008gravitational}, where it is shown that on a very long time scale $t\gg r$ the GW amplitude
decays, thus not leaving a permanent displacement characteristic of
a memory term. However the such test-particle displacement after the passage of
the GWs, even it not permanent, is present for $t\gtrsim r$, hence it can well
be seen as ``permanent'' on the human observation time scale when $r$ is an
astronomical scale.}

More recently the issue of memory effects was revisited in
\cite{Tolish:2014bka,Tolish:2014oda} where it has been shown that in the case
of a source involving the emission of a massless particle there is a memory
effect of the ``Christodoulou'' type, and in the toy model case of decay of
initial source into two massive particles the standard ``linear'' memory effect
is recovered.

However we stress once more that the effect we are presenting here is derived
from the \emph{non-radiative} part of the gravitational field, hence it is
not of the same type of various memory effects studied in literature, which
infer memory effects from the radiative part of the metric, see e.g. the recent
derivation of a ``velocity'' memory in plane gravitational waves
\cite{Zhang:2017rno,Zhang:2018srn}.
We will find in this work that the \emph{non-radiative} part of the metric
contributes to the Riemann tensor in eq.~(\ref{eq:geo_dev}) with a $\delta(t-r)/r^2$
term, being $\delta(t)$ an approximate Dirac delta function,
that does not generate a displacement memory\footnote{
  A term proportional to a $\delta$-function in the Riemann tensor
does not generate a displacement memory effect that would instead require a
$\dot\delta(t-r)/r$ term in the Riemann, denoting time derivative with an overdot.},
but rather a \emph{velocity} drift of test particles after the perturbation
has gone through them.

The outline of this paper is as follows. In sec.~\ref{sec:method} we present
the toy model for the source of gravitational field and how the linearised 
Einstein equations are going to be solved, reviewing the Scalar-Vector-Tensor
(SVT) formalism developed in \cite{Flanagan:2005yc}.
In sec.~\ref{sec:results} we show
the results of our analysis, and in sec.~\ref{sec:discussion} we interpret them
and investigate the detectability of the non-radiative
gravitational modes. We finally conclude in sec.~\ref{sec:conclusion}.

\section{Method}
\label{sec:method}
To consider the asymptotic behaviour of metric component in linearised General
Relativity, we find convenient to split the metric according to the SVT
decomposition introduced in \cite{Flanagan:2005yc}.
After modelling the source with an astrophysically realistic dynamics of emission
of particles/radiation, the metric components are obtained by solving
linarized Einstein equations. Then we concentrate our analysis on the
scalar, non-radiative degrees of freedom of the metric, looking for
effects present even after the passage of radiation and their observational
signature and observability will be studied in sec.~\ref{sec:discussion}.

According to the standard SVT decomposition, we parametrize the metric via
irreducible fields according to
\be
\ba{rl}
ds^2=&\ds-dt^2(1+2\phi)+2dt\,dx^i(\beta_i+b_{,i})\\
&\ds+dx^idx^j\paq{\delta_{ij}(1-2\psi)+\frac 12\pa{F_{i,j}+F_{j,i}}
+\pa{\dpa_i\dpa_j-\frac{\delta_{ij}}3\nabla^2}e+h_{ij}^{TT}}\,,
\ea
\ee
with constraints
\be
\ba{rcl}
\beta_{i,i}&=&0\,,\\
F_{i,i}&=&0\,,\\
h_{ii}^{TT}&=&0=h_{ij,j}^{TT}\,,
\ea
\ee
so that the fields $\phi,\psi,b,e,\beta_i,F_i,h_{ij}^{TT}$ have 10 independent
components in total parametrizing the 10 metric components.
Note that the constraints imply that the irreducible fields above are non-local
combinations of the original metric components $g_{\mu\nu}$, but the tidal Riemann fields can be written in terms of irreducible fields derivatives, which 
are local expression of the derivative of the original metric.
 
Consistently we apply the SVT decomposition to source components as well to obtain
\be
\label{eq:tmunu_svt}
\ba{rcl}
T_{00}&=&\ds \rho\,,\\
T_{0i}&=&\ds S_i+\dpa_iS\,,\\
T_{ij}&=&\ds p\delta_{ij}+\sigma_{ij}+\frac 12\pa{\sigma_{i,j}+\sigma_{j,i}}+
\pa{\dpa_i\dpa_j-\frac 13\delta_{ij}\nabla^2}\sigma\,,
\ea
\ee
parameterising the 10 degrees of freedom with the variables $\rho,S,S_i,p,\sigma,\sigma_i,\sigma_{ij}$, which, once the 6 constraints
$S_{i,i}=0=\sigma_{i,i}=\sigma_{ii}=\sigma_{ij,j}$ are applied, amount to 10
degrees of freedom.
Using now the energy momentum conservation, with $T_{0\mu}^{\ \ ,\mu}=0$ one has
\be
\label{eq:cons00}
\dot \rho =\nabla^2 S\implies S=\nabla^{-2}\dot\rho\,,
\ee
where the inverse Laplacian $\nabla^{-2}$ is defined to act on a test function
$f(t,\vec x)$ as
\be
\nabla^{-2}f(t,\vec x)=-\frac 1{4\pi}\int d^3x' \frac{f(t,\vec x-\vec x')}{|\vec x-\vec x'|}\,,
\ee
i.e. solutions to the Poisson equation are obtained from the source via the
standard Green's function of the Laplacian operator.
From $T_{i\mu}^{\ \ ,\mu}=0$ one obtains
\be
\label{eq:cons0i}
\dot S_i+\dpa_i\dot S&=&
p_{,i}+\frac 12\nabla^2\sigma_i+\frac 23\nabla^2\sigma_{,i}\nonumber\\
\implies&&\left\{
\ba{rclr}
\ds p+\frac 23\nabla^2\sigma&=&\ds\dot S=\nabla^{-2}\ddot \rho\,,\\
\sigma_i&=&\ds 2\nabla^{-2}\dot S_i
\ea\right.\,.
\ee
Following a standard procedure, see app.~\ref{app:svt} for details, one can
define the gauge invariant SVT-decomposed metric components
\be
\label{eq:giSVT}
\ba{rcl}
\ds\Phi&\equiv&\ds\phi+\dot b-\frac{\ddot e}2\,,\\
\ds\psi_e&\ds\equiv&\psi+\frac 16\nabla^2e\,,\\
\ds V_i&\equiv&\ds\beta_i-\frac 12\dot F_i\,,\\
\ds h_{ij}^{TT}&\,,&
\ea
\ee
and obtain the following expressions for the Einstein equations:
\be
\label{eq:cov}
\ba{rcll}
\nabla^2\psi_e&=&\ds 4\pi G_N\rho\,,& \rm{S:}\,00\,,0i\,,\\
\nabla^2\Phi&=&\ds 4\pi G_N\pa{\rho+3p-3\dot S}\,,&\rm{S:}\,ij\,,\\
\nabla^2V_i&=&\ds -16\pi G_N S_i\,,&\rm{V:}\,0i\,,ij\,,\\
\Box h_{ij}^{TT}&=&-16\pi G_N\sigma_{ij}\,,&\rm{T:}\,ij\,,
\ea
\ee
Although the gauge invariant variables $\Phi, \psi_e, V_i, h_{ij}^{TT}$ are non-local
quantities of the metric component $g_{\mu\nu}$,
they appear in a local combinations in the Riemann components, e.g. at linear order
\be
R^i_{\ 0j0}=\delta_{ij}\ddot\psi_e+\Phi_{,ij}+\frac 12\pa{\dot V_{i,j}+\dot V_{j,i}}-
\frac 12\ddot h_{ij}^{TT}\,.
\ee
As long as observables depend on Riemann tensor components, they can be expressed
indifferently in terms of $g_{\mu\nu}$ or $\psi_e,\Phi,V_i,h_{ij}^{TT}$.

The physical situation we are going to consider corresponds to a source whose
constituents change suddenly from a static object following a time-like geodesic
to an expanding shell of radiation.
Such system arises naturally in a variety of astrophysical situation, hence
represent a training ground of phenomenological interest to test how memory
effects can be recorded in the non-radiative part of the metric perturbations.

\section{Results}
\label{sec:results}
To illustrate our result we consider as a source a massive object at rest at
the origin of coordinates for $t<0$ (term proportional to $1-\Theta(t)$)
which turns into an expanding shell of radiation for $t>0$
(term proportional to $\Delta(t-r)$),
with $\Theta(t)$ denoting a generic monotonic
smooth function interpolating between 0 for
$t\to -\infty$ to 1 for $t\to\infty$, whose main variation occurs around
$t\sim 0$, and $\Delta(x)\equiv \dfrac{{\rm d}\Theta(x)}{{\rm d}x}$.
The functions $\Theta(x)$ and $\Delta(x)$ can be seen respectively
as approximate Heaviside step and Dirac delta functions.

This example is simple enough to be treated analytically and at the
same time, as it will be shown, it contains the hearth of the physical result
we are going to investigate.
The energy density of the system can be quantitatively expressed as
\be
T_{00}(t,\vec x)=\pa{1-\Theta(t)}M\delta^{(3)}(\vec x)+
\frac M{4\pi r^2}\Delta(t-r)\,,
\ee
with $r=|\vec x|$.
The presence of isotropic radiation leads to assume an isotropic pressure given
by
\be
p=\frac M{12\pi r^2}\Delta(t-r)\,,
\ee
so that the scalar components of the energy-momentum tensor are completely
specified: the scalar part of eq.~(\ref{eq:cons0i}), which is the
continuity equation for the source current
($S_i$ drops out of the continuity equation being divergence-free), is solved by
\be
S=\ds\frac M{4\pi}\int_r^\infty dr'\frac{\Delta(t-r')}{{r'}^2}\,.
\ee
Note that 
See app.~\ref{app:pots} for detailed derivations of $S,\sigma$ and the scalar
metric fields, which can be written as\footnote{A similar setup was studied in \cite{Garfinkle:2014rpa}
  where a potential analog to our $\psi_e$ was erroneously found to be
  time-independent, by first assuming that $t/r\ll 1$ and then taking the limit
  $t\to \infty$, see eqs. (22), (24) and (25) there.}
\be
\label{eq:psi_phi_iso}
\ba{rcl}
\ds\psi_e&=&\ds -\paq{1-\Theta(t-r)}\frac{G_NM}{r}
-G_NM\int_r^\infty dr'\frac{\Delta(t-r')}{r'}\,,\\
\ds\Phi&=&\ds-\paq{1-\Theta(t-r)}\frac{G_NM}r
-G_NM\int_r^\infty dr'\frac{\Delta(t-r')}{r'}\pa{2-\frac{r^2}{{r'}^2}}\,.
\ea
\ee
Note that while the source undergoes a discontinuity at $t=0$, fields
$\Phi,\psi_e$ at the location of the observer are constant until $t\sim r$,
i.e. when radiation pass through the observer, as required by causality.

As expected, $\psi_e$ and $\Phi$ are continuous at $t=0$. For $\psi_e$
this is trivial since it can be obtained by
convolving the Green function with $\rho$, which is continuous function of time.
The combination
$\phi-\psi_e$ is proportional to the convolution of the Green function with
$p-\dot S$, and while both $p$ and $\dot S$ individually change discontinuously
at $t=0$ for $r=0$, their combined contribution to the potential is also
constant around $t=0$, see app.~\ref{app:pots} for detailed calculations, thus
leading to a continous $\Phi$.

In particular their early/late time behaviour of $\psi_e$
\be
\ba{rcl}
\psi_e(t\ll r)&=&\ds -\frac{G_NM}r\,,\\
\psi_e(t\gg r)&=&\ds -\frac{G_NM}t\,,
\ea
\ee
are the same as the Newtonian potential outside ($t\ll r$) or inside ($t\gg r$)
a spherical shell of matter with mass $M$. The potential $\Phi$ is equal to $\psi_e$ at early times, and one has $\Phi(t)-\psi_e(t)\simeq -G_NM/t$ for $t\gg r$.

The scalar part of the Riemann component entering the geodesic deviation equation reads, see again app.~\ref{app:pots} for derivation:
\be
\label{eq:Riem_sph}
\ba{rcl}
\ds\left.R^i_{\ 0j0}\right|_S&=&\ds\ddot \psi_e\delta^i_{\ j}+\Phi^{,i}_{\ j}\\
&=&\ds\frac{G_NM}{r^2}\paq{\pa{\delta^i_{\ j}-\frac{x^ix_j}{r^2}}\Delta(t-r)
  +\pa{\delta^i_{\ j}-3\frac{x^ix_j}{r^2}}\frac{\pa{1-\Theta(t-r)}}r}\,.
\ea
\ee

We first notice in the scalar part of the Riemann the presence of the standard
$1/r^3$ tidal terms for the static source for early enough time which vanishes
after the expanding sphere of radiation has passed through the test particles.

More interesting is the other term, proportional to $\Delta(t-r)/r^2$: after
integrating once over time the (scalar part of the) geodesic deviation equation
on a small time interval around $t\simeq r$ this term is responsible for a 
\emph{transverse velocity kick} to test particles, as
\be
\label{eq:vel_kick}
\Delta_\epsilon \dot D^i=\int_{r-\epsilon}^{r+\epsilon}\ddot D^i dt\simeq
\pa{\delta^i_j-\frac{x^ix_j}{r^2}}\frac{G_NM}{r^2}D^j\,,
\ee
where $\epsilon$ is the shortest time for which $\Delta(\epsilon)\sim 0$
so that $\int_{-\epsilon}^\epsilon\Delta(t)dt=1$.
Two test particles (with separation perpendicular to $\vec r$) at mutual rest
before the expanding spherical shell of radiation pass through them,
will acquire a velocity drift with respect to each other after the passage of
radiation, bearing the ``memory'' of such passage at arbitrarily late times.

Note that the Riemann tensor is completely causal, depending on the status 
of the source at retarded time $t-r$, the ``memory'' effects just appear
in the status of motion of test particles.

A \emph{displacement} memory type term would have been like
$G_NM\dot \Delta(t-r)/r$, able to give a displacement kick to test
particles after double integration over time around $t\simeq r$ of the geodesic
deviation equation (\ref{eq:geo_dev}), giving rise to $\Delta D^i\simeq G_NM/r D^i$.
A displacement kick $\propto 1/r$ is indeed originated from the
\emph{tensorial}, radiative
part of the geodesic deviation equation and it is the linear Christodoulou
effect.

Finally we observe that $R_{00}=\delta^j_iR^i_{\ 0j0}$ correctly vanishes outside
the sources.

\section{Discussion}
\label{sec:discussion}
To evaluate the phenomenological impact of the first term on the right hand side of eq.~(\ref{eq:Riem_sph})
one can integrate the geodesic deviation equation in a small time interval
around $t=r$ for two test masses at distance $D$, as done in
eq.~(\ref{eq:vel_kick}), to get
\be
\label{eq:vk}
\Delta \dot D^i=\pa{\delta^i_j-\frac{x^ix_j}{r^2}}\frac{G_NM}{c\, r^2}D^i\,,
\ee
where the speed of light $c$ has been re-instated.
Eq.~(\ref{eq:vk}) shows that two test masses with transverse separation $D$ to
with respect to the source-observer direction will experience a velocity drift
with magnitude $\sim G_NMD/(cr^2)$.

The detectability of a (Fourier-transformed) metric perturbation $\tilde h(f)$
by a detector with (single-sided) noise spectral density $S_n(f)$ is quantified
\footnote{The single-sided noise spectral density is defined in terms of the
  average of the Fourier transform of the dimension-less strains $\tilde n(f)$ via
$\langle \tilde n(f)\tilde n^*(f')\rangle=\frac 12S_n(f)\delta(f-f')$.}
by its Signal-to-Noise-Ratio (SNR), see e.g. the standard textbook
\cite{maggiore2008gravitational} at ch. 7:
\be
\label{eq:SNR}
{\rm SNR}^2=4\int_0^\infty d(\log f)\frac{f|\tilde h(f)|^2}{S_n(f)}\,,
\ee
where a threshold for detection of $SNR\geq 8$ is usually assumed.
From the second derivative of the strain responsible for the velocity
kick $\propto \Delta(t-r)$, using that $|\tilde h(f)|=|\tilde{\ddot h}/(2\pi f)^2|$,
one gets a velocity kick strain $\tilde h_{vk}(f)$
\be
\label{eq:hvk}
\tilde h_{vk}(f)\sim \frac{G_NM}{c\pa{2\pi fr}^2}\,,
\ee
from which one can estimate a dimension-less strain $h_c$ given by
$$
h_c=f\tilde h_{vk}\simeq 2\times 10^{-27}\pa{\frac M{M_\odot}}
\pa{\frac{f}{10^{-4}{\rm Hz}}}^{-1}\pa{\frac r{r_{GC}}}^{-2}\,,
$$
with $r_{GC}$ the solar system distance to the galactic center $r_{GC}\sim 8$ kpc.

\begin{figure}
  \begin{center}
    \includegraphics[width=.6\linewidth]{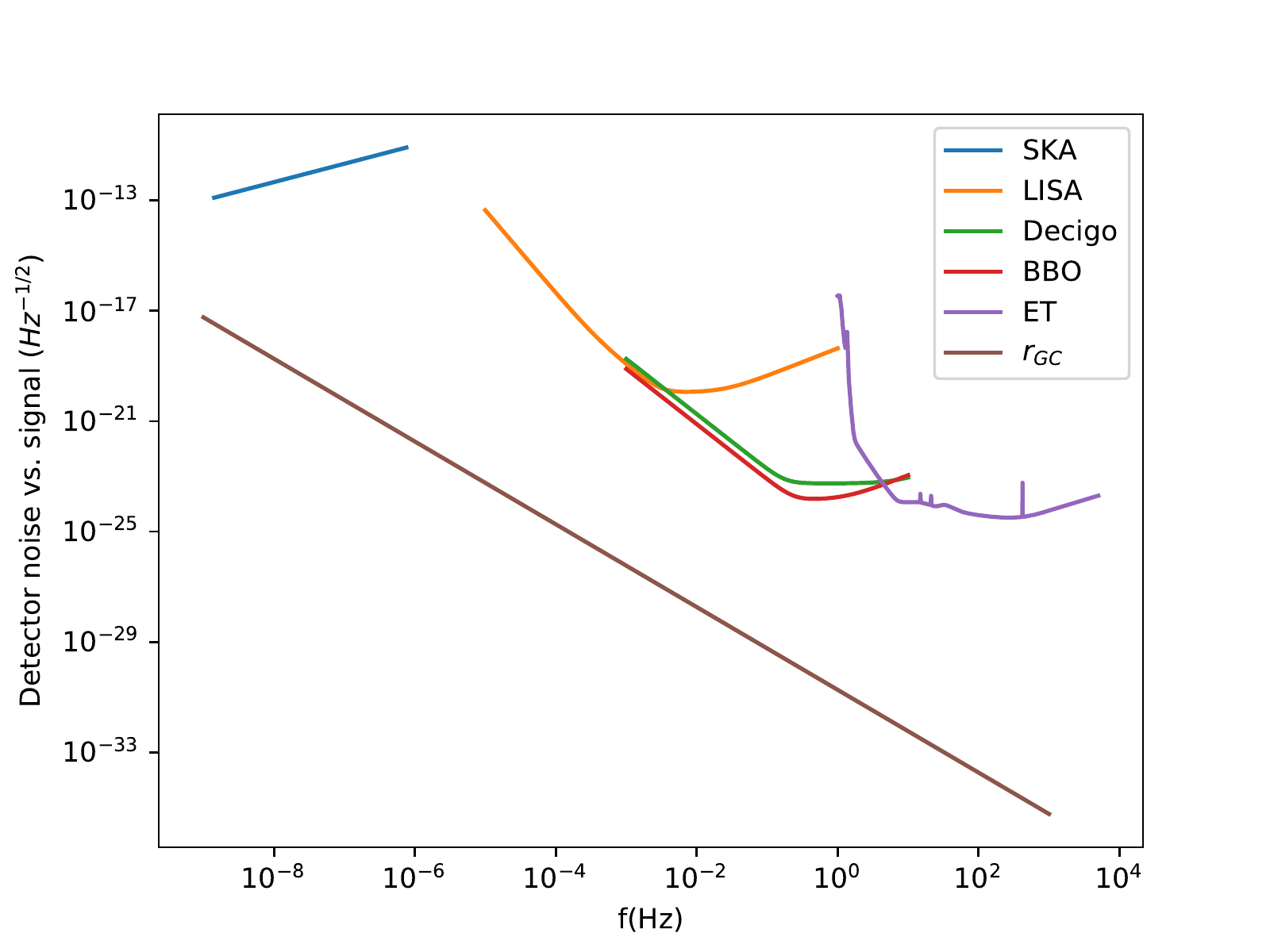}
    \caption{Square root of the noise spectral density of several planned GW
      detectors compared to the
      velocity-kick amplitude $\sqrt f|\tilde h_{vk}|$ for $M=M_\odot$ at
      $r=r_{GC}=8$ kpc, which is the approximate distance to the galactic center.
      Also shown for comparison spectral noise densities for different detectors:
      SKA \cite{Moore:2014lga}, LISA \cite{lisanoise}, DECIGO \cite{Kawamura:2011zz},
      BBO \cite{Cutler:2009qv}, ET \cite{Hild_2011bis}.}
  \label{fig:det_vk}
  \end{center}
\end{figure}

Considering the possibility to observe this effect by the future GW detectors,
fig.~\ref{fig:det_vk} shows the square root of $S_n(f)$ (i.e. the denominator in the SNR
integrand in eq.~(\ref{eq:SNR})) for various future GW detectors, and
$\sqrt f |\tilde h_{vk}(f)|$ (i.e. the square root of the numerator of the SNR integrand), considering an optimistic mass release $M=M_\odot$ for a source
located at the galactic center.

The reason for the smallness of the effect is that it arises from a \emph{longitudinal}
mode, affecting the Riemann curvature as the inverse \emph{square} of the distance:
from the geodesic deviation of test particles at relative separation $\Delta L$ the velocity
memory effect
is of the type $\Delta v \sim \dot h \Delta L$, being $h$ the metric perturbation.
For the longitudinal metric perturbation considered in this work the amplitude
of the transient
$\dot h$ giving rise to the memory effect is $\sim GM/r^2$,
whereas a transient radiative mode can give rise to a velocity
memory with $\dot h\sim \omega h$, being $\omega$ the typical oscillation
frequency of the gravitational wave, see e.g. \cite{Zhang:2018srn}.

\section{Conclusion}
\label{sec:conclusion}
By analysing a toy model of emission of generic radiation by a compact
astrophysical object we find simple analytic expressions for the time-dependent
scalar metric potentials and the scalar part of the Riemann curvature.

We find that the scalar potentials and curvature at the observer site undergo
variations only at the time when the perturbations reach the observers, as
expected by causality. Beside the standard $1/r^3$ almost-static tidal term,
the scalar part of the curvature contains a term producing a ``step function''
in the relative velocity of test masses between immediately before and after
the passage of radiation through detector test masses, like mirrors in
gravitational observatories.
The resulting \emph{velocity kick} amplitude varies with the inverse of the
\emph{square} of the source-detector distance.

From the observational point of view the dependence on the inverse square
distance, in sharp contrast with the inverse distance dependence of the
gravitational waves, suppresses such effect making virtually impossible to be
detect with GW detectors.

\section*{Acknowledgements}
The authors wishes to thank Jan Harms and Robert Wald for discussions.
This study was financed in part by the Coordena\c{c}\~ao de Aperfei\c{c}oamento de
Pessoal de N\'\i vel Superior - Brasil (CAPES) - Finance Code 001.
The work of RS is partially supported by CNPq.

\appendix

\section{Einstein equations in the SVT decomposition}
\label{app:svt}
Defining $\psi_e\equiv\psi+1/6\nabla^2 e$ and $V_i\equiv \beta_i-1/2\dot F_i$,
one has for the Christoffel coefficients
\be
\ba{rcl}
\Gamma^0_{00}&=&\dot\phi\,,\\
\Gamma^0_{0i}&=&\phi_{,i}\,,\\
\Gamma^0_{ij}&=&\ds-\delta_{ij}\dot\psi_e+\frac 12\dot e_{,ij}-b_{,ij}
-\frac 12\pa{V_{i,j}+V_{j,i}}+\frac 12 \dot h^{TT}_{ji}\,,\\
\Gamma^k_{00}&=&\ds\delta^{km}\pa{\phi_{,m}+\dot b_{,m}+\dot \beta_m}\,,\\
\Gamma^k_{0i}&=&\ds-\delta^k_i\dot\psi_e+\frac 12\delta^{km}\dot e_{,im}
+\frac 12\pa{V_{k,i}-V_{i,k}}+\frac 12\delta^{km}\dot h^{TT}_{mi}\,,\\
\Gamma^k_{ij}&=&\ds\delta^{km}\delta_{ij}\psi_{e,m}-\delta^k_i\psi_{e,j}-\delta^k_j\psi_{e,i}+\frac 12\delta^{kl}e_{,ijk}
+\frac 12\delta^{km}F_{l,im}+\frac 12\delta^{km}\pa{h^{TT}_{im,j}+h^{TT}_{jm,i}-h^{TT}_{ij,m}}\,.
\ea
\ee
and for Riemann tensor components
\be
\ba{rcl}
R^0_{\ i0j}&=&\ds-\delta_{ij}\ddot\psi_e-\pa{\phi+\dot b-\frac 12\ddot e}_{,ij}
-\frac 12\pa{\dot V_{i,j}+\dot V_{j,i}}+\frac 12\ddot h^{TT}_{ij}\,,\\
R^k_{\ ilj}&=&\ds\delta^{km}\pa{\delta_{ij}\psi_{e,ml}-\delta_{il}\psi_{e,mj}}
+\delta^k_l\psi_{e,ij}-\delta^k_j\psi_{e,il}
+\frac 12 \delta^{km}\pa{h^{TT}_{jm,il}+h^{TT}_{il,jm}-h^{TT}_{ij,lm}-h^{TT}_{lm,ij}}\,,\\
R^0_{\ ijk}&=&\ds\delta_{ij}\dot\psi_{e,k}-\delta_{ik}\dot\psi_{e,l}
+\frac12\pa{V_{j,ik}-V_{k,ij}}+\frac 12\pa{\dot h^{TT}_{ik,j}-\dot h^{TT}_{ij,k}}\,.
\ea
\ee
Now defining $\Phi\equiv \phi+\dot b-\ddot e/2$ one has
\be
\ba{rcl}
R_{00}&=&\ds 3\ddot \psi_e+\nabla^2\Phi\,,\\
R_{0i}&=&\ds 2\psi_{e,i}-\frac 12\nabla^2 V_i\,,\\
R_{ij}&=&\ds \delta_{ij}\pa{-\ddot \psi_e+\nabla^2\psi_e}+\psi_{e,ij}-\Phi_{,ij}-\frac 12\pa{\dot V_{i,j}+\dot V_{j,i}}+\frac 12\ddot h^{TT}_{ij}-\frac 12\nabla^2 h^{TT}_{ij}\,,
\ea
\ee
hence
\be
R=-6\ddot \psi_e-2\nabla^2\Phi+4\nabla^2\psi_e\,.
\ee
The Einstein's equations can be split according to their representation under
rotation:
\begin{itemize}
\item Scalar
\be
\ba{rcl}
G_{00}&=&\ds 2\nabla^2\psi_e\,,\\
G_{0i}&=&\ds 2\dot\psi_{e,i}\,,\\
G_{ij}&=&\delta_{ij}\paq{2\ddot \psi_e+\nabla^2(\Phi-\psi_e)}+\pa{\psi_e-\Phi}_{,ij}\,.
\ea
\ee
\item Vector
\be
\ba{rcl}
G_{00}&=&0\,,\\
G_{0i}&=&\ds-\frac 12\nabla^2 V_i\,,\\
G_{ij}&=&\ds-\frac 12\paq{\dot V_{i,j}+\dot V_{j,i}}\,.
\ea
\ee
\item Tensor
\be
\ba{rcl}
G_{00}&=&\ds G_{0i}=0\,,\\
G_{ij}&=&\ds\frac 12\ddot h^{TT}_{ij}-\frac 12\nabla^2h^{TT}_{ij}\,,
\ea
\ee
\end{itemize}
The action of a coordinate transformation
\be
\ba{rclcl}
x^0&\to& {x'}^0&=&x^0+\xi^0\,,\\
x^i&\to& {x'}^i&=&x^i+\xi^i+a^{,i}\,,
\ea
\ee
with $\xi^i_{,i}=0$, has the following effects on SVT fields:
\be
\ba{rclcl}
\phi&\to&\phi'&=&\phi-\dot\xi^0\,,\\
b&\to& b'&=&b+\xi^0-\dot a\,,\\
\beta_i&\to&\beta_i'&=&\beta_i-\dot\xi^i\,,\\
\psi&\to&\psi'&=&\psi+\frac 13\nabla^2a\,.\\
F_i&\to& F_i'&=&F_i-2\xi_i\,,\\
e&\to& e'&=&e-2a\,,
\ea
\ee
from which one can verify that fields in eq.~(\ref{eq:giSVT}) are coordinate
transformation invariant, at linear order.

\section{Detailed derivation of scalar potentials}
\label{app:pots}
The inverse Laplacian operator appears in several equations applied to
spherically symmetric functions, and it involves integrals of the type
\be
-\frac 1{4\pi}\int d^3x'\frac 1{|\vec x-\vec x'|}f(r)=
-\frac 1r\int_0^r f(r'){r'}^2dr'-\int_r^\infty f(r')r'dr'\,,
\ee
where it has been used that
\be
\frac 1{|\vec r(\theta,\phi)-\vec r(\iota,\alpha)|}=\frac {4\pi}{r_>}
\sum_{l\geq 0}\sum_{|m|\leq l}\frac 1{2l+1}\pa{\frac{r_<}{r_>}}^l
Y_{lm}^*(\theta,\phi)Y_{lm}(\iota,\alpha)\,,
\ee
where $r\equiv |\vec r(\theta,\phi)|$, $r'\equiv |\vec r'(\iota,\alpha)|$,
$r_{<,>}\equiv max,min(r,r')$ and spherical symmetry has been used.

We complement the results of sec.~\ref{sec:results} by providing the explicit
derivation of $S$ from the first of eqs.(\ref{eq:cons0i}):
\be
\ba{rcl}
\ds S&=&\ds-\frac 1{4\pi}\int d^3 x\frac{\dot \rho}{|\vec x-\vec x'|}\\
&=&\ds \frac{\Delta(t)}{4\pi}\frac Mr-\frac M{4\pi}\pa{
  \frac 1r\int_0^r \dot\Delta(t-r')dr'+
  \int_r^\infty \frac{\dot \Delta(t-r')}{r'} dr'}\\
&=&\ds \frac M{4\pi}\int_r^\infty\frac{\Delta(t-r')}{{r'}^2}dr'\,,
\ea
\ee
where integration by parts and the relation
$\dot\Delta(t-r)=-\frac{{\rm d}\Delta(t-r)}{{\rm d}r}$ have been used.

Analogously $\sigma$, the remaining scalar component of the energy-momentum tensor,
can be found from the second of eqs.~(\ref{eq:cons0i}) (we remind that
$\dot \Theta(t)=\Delta(t)$)
\be
\ba{rcl}
\sigma&=&\ds \frac M{32\pi^2}\int d^3x'\frac 1{|\vec x-\vec x'|}
\pa{\frac{\Delta(t-r')}{{r'}^2}-3\int_{r'}^\infty\frac{\dot\Delta(t-r'')}{{r''}^2}dr''}\\
&=&\ds \frac M{4\pi}\left[
  -\frac 1r\int_0^r\Delta(t-r')dr'-\int_r^\infty \frac{\Delta(t-r')}{r'}dr'
  \right.\\
  &&\ds\left.\qquad+\frac 3r\int_0^r dr' {r'}^2\int_{r'}^\infty \frac{\Delta(t-r'')}{{r''}^3}dr''
  +3\int_r^\infty dr' r'\int_{r'}^\infty \frac{\Delta(t-r'')}{{r''}^3}dr''
  \right]\\
&=&\ds \frac{M}{4\pi}\left[
  -\frac 1r\int_0^r\Delta(t-r')dr'-\int_r^\infty \frac{\Delta(t-r')}{r'}dr'+
  \frac 3r\int_0^rdr''\frac{\Delta(t-r'')}{{r''}^3}\int_0^{r''}{r'}^2dr'
  \right.\\
  &&\ds\qquad\left.
  +3\int_r^\infty dr''\frac{\Delta(t-r'')}{{r''}^3}
  \pa{\frac 1r\int_0^r{r'}^2dr'+\int_r^{r''}r'dr'}\right]\\
&=&\ds \frac M{8\pi}\int_r^\infty\frac{\Delta(t-r')}{r'}
\pa{1-\frac{r^2}{{r'}^2}}dr'\,,
\ea
\ee
where integration by parts has been used in the first passage and the
integration order over $r',r''$ has been exchanged in the following one
(and in the final passage $r''$ has been substitud with $r'$).

The solution for the scalar potential $\psi_e$ can be found by direct
integration of the first of eqs.~(\ref{eq:cov})
\be
\label{eq:psie}
\ba{rcl}
\psi_e&=&\ds -G_N\int d^3x\frac \rho {|\vec x-\vec x'|}\\
&=&\ds-\paq{1-\Theta(t)}\frac{G_NM}r-G_NM\pa{
  \frac 1r\int_0^r\Delta(t-r')dr'+\int_r^\infty\frac{\Delta(t-r')}{r'}dr'}\\
&=&\ds-\paq{1-\Theta(t-r)}\frac{G_NM}r-G_NM\int_r^\infty\frac{\Delta(t-r')}{r'}dr'\\
&=&\ds-\frac{G_NM}r+G_NM\int_r^\infty\frac{\Theta(t-r')}{{r'}^2}dr'\,,
\ea
\ee
where integrations by parts and $\Delta(t-r)=-\frac{d\Theta(t-r)}{dr}$ have
been used repeatedly.

Taking the difference of the first two of eqs.~(\ref{eq:cov}) and noting that
from the first of eqs.(\ref{eq:cons0i}) one has $2\nabla^2\sigma=3(\dot S-p)$,
one finds $\psi_e-\Phi=8\pi G_N\sigma$, which using eq.(\ref{eq:psie}) allows
to derive
\be
\label{eq:Phi}
\ba{rcl}
\Phi&=&\ds -G_NM\int_r^\infty\frac{\Delta(t-r')}{r'}\pa{1-\frac{r^2}{{r'}^2}}dr'-\frac{G_NM}r+G_NM\int_r^\infty\frac{\Theta(t-r')}{{r'}^2}dr'\\
&=&\ds -\frac{G_NM}r+G_NM\int_r^\infty\frac{\Theta(t-r')}{{r'}^2}\pa{2-3\frac{r^2}{{r'}^2}}\\
&=&\ds -\paq{1-\Theta(t-r)}\frac{G_NM}r-G_NM\int_r^\infty\frac{\Delta(t-r')}{r'}\pa{2-\frac{r^2}{{r'}^2}}\,.
\ea
\ee

To compute the scalar part of the Riemann tensor entering the
geodesic deviation equations one still needs 
\be
\label{eq:ddpsi}
\ba{rcl}
\ds\ddot\psi_e&=&\ds G_NM\int_r^\infty\frac{\dot\Delta(t-r)}{{r'}^2}dr'\\
&=&\ds \Delta(t-r)\frac{G_NM}{r^2}-2G_NM\int_r^\infty\frac{\Delta(t-r')}{{r'}^3}dr'\,,
\ea
\ee
and the gradients of $\Phi$, which starting from the next to last line in
eq.~(\ref{eq:Phi}) can be written as
\be
\label{eq:grad_phi}
\ba{rcl}
\ds\Phi_{,i}&=&\ds x_i\pag{\paq{1+\Theta(t-r)}\frac{G_NM}{r^3}
  -6G_NM\int_r^\infty\frac{\Theta(t-r')}{{r'}^4}dr'}\,,\\
\ds \Phi_{,ij} &=&\ds \delta_{ij}
\pag{\paq{1+\Theta(t-r)}\frac{G_NM}{r^3}
  -6G_NM\int_r^\infty\frac{\Theta(t-r')}{{r'}^4}dr'}\\
&&\ds+\frac{x_ix_j}{r^2}\left\{-\Delta(t-r)\frac{G_NM}{r^2}
  -3\paq{1+\Theta(t-r)}\frac{G_NM}{r^3}+6\Theta(t-r)\frac{G_NM}{r^3}\right\}\\
  &=&\ds \delta_{ij}\pag{\paq{1-\Theta(t-r)}\frac{G_NM}{r^3}
    +2G_NM\int_r^\infty\frac{\Delta(t-r')}{{r'}^3}dr'}\\
  &&\ds+\frac{x_ix_j}{r^2}\pag{-\Delta(t-r)\frac{G_NM}{r^2}
  -3\paq{1-\Theta(t-r)}\frac{G_NM}{r^3}}\,.
\ea
\ee
Finally combining eqs.(\ref{eq:ddpsi}) and (\ref{eq:grad_phi}) one gets eq.~(\ref{eq:Riem_sph}).

\bibliographystyle{spphys}

\begin{thebibliography}{10}
\providecommand{\url}[1]{{#1}}
\providecommand{\urlprefix}{URL }
\expandafter\ifx\csname urlstyle\endcsname\relax
  \providecommand{\doi}[1]{DOI \discretionary{}{}{}#1}\else
  \providecommand{\doi}{DOI \discretionary{}{}{}\begingroup
  \urlstyle{rm}\Url}\fi

\bibitem{LIGOScientific:2018mvr}
B.~Abbott, et~al., Phys. Rev. X \textbf{9}(3), 031040 (2019).
\newblock \doi{10.1103/PhysRevX.9.031040}

\bibitem{Abbott:2020niy}
R.~Abbott, et~al., arXiv:2010.14527

\bibitem{1974SvA....18...17Z}
Y.B. {Zel'dovich}, A.G. {Polnarev}, Sov. Ast. \textbf{18}, 17 (1974)

\bibitem{Braginsky:1986ia}
V.B. Braginsky, L.P. Grishchuk, Sov. Phys. JETP \textbf{62}, 427 (1985)

\bibitem{Thorne:1992sdb}
K.S. Thorne, Phys. Rev. \textbf{D45}(2), 520 (1992).
\newblock \doi{10.1103/PhysRevD.45.520}

\bibitem{Wiseman:1991ss}
A.G. Wiseman, C.M. Will, Phys. Rev. \textbf{D44}(10), R2945 (1991).
\newblock \doi{10.1103/PhysRevD.44.R2945}

\bibitem{Christodoulou:1991cr}
D.~Christodoulou, Phys. Rev. Lett. \textbf{67}, 1486 (1991).
\newblock \doi{10.1103/PhysRevLett.67.1486}

\bibitem{1978ApJ...223.1037E}
R.~{Epstein}, Astrophysical Journal \textbf{223}, 1037 (1978).
\newblock \doi{10.1086/156337}

\bibitem{Turner:1978jj}
M.S. Turner, Nature \textbf{274}, 565 (1978).
\newblock \doi{10.1038/274565a0}

\bibitem{1997A&A...317..140M}
E.~{Mueller}, H.T. {Janka}, Astron. Astrophys. \textbf{317}, 140 (1997)

\bibitem{Blanchet:1987wq}
L.~Blanchet, T.~Damour, Phys. Rev. \textbf{D37}, 1410 (1988).
\newblock \doi{10.1103/PhysRevD.37.1410}

\bibitem{Blanchet:1992br}
L.~Blanchet, T.~Damour, Phys. Rev. \textbf{D46}, 4304 (1992).
\newblock \doi{10.1103/PhysRevD.46.4304}

\bibitem{maggiore2008gravitational}
M.~Maggiore, \emph{Gravitational Waves: Volume 1: Theory and Experiments}.
\newblock Gravitational Waves (OUP Oxford, 2008).
\newblock \urlprefix\url{https://books.google.com.br/books?id=AqVpQgAACAAJ}

\bibitem{Tolish:2014bka}
A.~Tolish, R.M. Wald, Phys. Rev. \textbf{D89}(6), 064008 (2014).
\newblock \doi{10.1103/PhysRevD.89.064008}

\bibitem{Tolish:2014oda}
A.~Tolish, L.~Bieri, D.~Garfinkle, R.M. Wald, Phys. Rev. \textbf{D90}(4),
  044060 (2014).
\newblock \doi{10.1103/PhysRevD.90.044060}

\bibitem{Zhang:2017rno}
P.M. Zhang, C.~Duval, G.~Gibbons, P.~Horvathy, Phys. Lett. B \textbf{772}, 743
  (2017).
\newblock \doi{10.1016/j.physletb.2017.07.050}

\bibitem{Zhang:2018srn}
P.~Zhang, C.~Duval, G.~Gibbons, P.~Horvathy, JCAP \textbf{05}, 030 (2018).
\newblock \doi{10.1088/1475-7516/2018/05/030}

\bibitem{Flanagan:2005yc}
E.E. Flanagan, S.A. Hughes, New J. Phys. \textbf{7}, 204 (2005).
\newblock \doi{10.1088/1367-2630/7/1/204}

\bibitem{Garfinkle:2014rpa}
D.~Garfinkle, I.~Ra\'acz, Gen. Rel. Grav. \textbf{47}(7), 79 (2015).
\newblock \doi{10.1007/s10714-015-1924-2}

\bibitem{Moore:2014lga}
C.~Moore, R.~Cole, C.~Berry, Class. Quant. Grav. \textbf{32}(1), 015014 (2015).
\newblock \doi{10.1088/0264-9381/32/1/015014}

\bibitem{lisanoise}
L.S. study team, Lisa science requirements document.
\newblock Tech. rep., ESA (2018).
\newblock Https://www.cosmos.esa.int/web/lisa/lisa-documents, accessed on Aug
  27th, 2020

\bibitem{Kawamura:2011zz}
S.~Kawamura, et~al., Class. Quant. Grav. \textbf{28}, 094011 (2011).
\newblock \doi{10.1088/0264-9381/28/9/094011}

\bibitem{Cutler:2009qv}
C.~Cutler, D.E. Holz, Phys. Rev. D \textbf{80}, 104009 (2009).
\newblock \doi{10.1103/PhysRevD.80.104009}

\bibitem{Hild_2011bis}
S.H. et~al., Classical and Quantum Gravity \textbf{28}(9), 094013 (2011).
\newblock \doi{10.1088/0264-9381/28/9/094013}.
\newblock
  \urlprefix\url{https://doi.org/10.1088\%2F0264-9381\%2F28\%2F9\%2F094013}

\end{thebibliography}

\end{document}